# The pseudogap in $Bi_2Sr_2CaCu_2O_{8+x}$ single crystals from tunneling measurements: a possible evidence for the Cooper pairs above $T_c$.

A. Mourachkine

*Université Libre de Bruxelles, Service de Physique des Solides, CP233, Blvd du Triomphe, B-1050 Brussels, Belgium*

**Abstract.** - We present electron-tunneling spectroscopy of slightly overdoped $Bi_2Sr_2CaCu_2O_{8-x}$ (Bi2212) single crystals with $T_c$ = 87 - 90 K in a temperature range between 14 K and 290 K using a break-junction technique. The pseudogap which has been detected above $T_c$ appears at $T^*$ 280 K. The analysis of the spectra shows that there is a contribution to the pseudogap above $T_c$, which disappears approximately at 110 - 115 K. We associate this contribution with the presence of incoherent Cooper pairs.

The existence of a pseudogap in electronic excitation spectra of high-$T_c$ superconductors (HTSC), which appears below a certain temperature $T^* > T_c$, is considered to be amongst the most important features of cuprates. Many experiments have provided evidence (NMR [1], angle-resolved photoemission (ARPES) [2], specific heat [3], electron-tunneling spectroscopy [4] and STM [5]) for a gap-like structure in electronic excitation spectra. There is a consensus on that the pseudogap is a characteristic feature of the underdoped regime of copper-oxides, and the value of $T^*$ increases with the decrease of hole concentration. For example, in tunneling measurements on underdoped Bi2212 single crystals by STM, the pseudogap was detected even at room temperature [5]. At the same time, tunneling measurements showed that the pseudogap is also present in slightly overdoped samples as well [4]. An advantage of tunneling spectroscopy is that it is particularly sensitive to the density of state (DOS) near the Fermi level $E_F$ and, thus, is capable of detecting any gap in the quasiparticle excitation spectrum at $E_F$ [6]. In addition to this, tunneling spectroscopy has a high energy resolution [6].

In general, the pseudogap in cuprates can originate from the charge-density-wave (CDW) order, the spin-density-wave (SDW) order due to local antiferromagnetic (AF) correlations, the presence of incoherent Cooper pairs, or from their combination (so-called *Ansatz*) [7]. From earlier studies, the pseudogap was associated with a SDW gap due to local AF correlations [1]. Very recent measurements of high-frequency conductivity that track the phase-correlation time in the normal state of underdoped Bi2212 showed that the incoherent Cooper pairs remain up to $T_{\text{pair}}$ which is well above $T_c$, however, lower than $T^*$ [8]. Thus, between $T_c$ and $T_{\text{pair}}$, there is a contribution of incoherent Cooper pairs to the pseudogap. Indeed, there is a consensus on that, to the contrary to conventional SCs, the formation of the Copper pairs and the establishment of the phase coherence among them in cuprates occur at different temperatures, $T_{\text{pair}} > T_c$ [9,8,5]. So, it is reasonable to expect that the incoherent Cooper pairs have to be observed above $T_c$ in different types of measurements including tunneling. The $T_{\text{pair}}$ scales with $T^*$, thus, the difference between $T_{\text{pair}}$ and $T_c$ is the highest in the underdoped regime and becomes smaller in the overdoped regime [9].

We present here direct measurements of the density of states by electron-tunneling spectroscopy on slightly overdoped Bi2212 single crystals using a break-junction technique. The pseudogap which has been detected above $T_c$ appears at $T^*$ 280 K. The analysis of the spectra shows that there is a contribution to the pseudogap above $T_c$, which disappears between $T_c$ and $T^*$. We associate this contribution to the pseudogap with the presence of incoherent Cooper pairs in Bi2212. To our knowledge, the data shown in Fig. 2 are the most detailed *tunneling* data of the pseudogap presented in the literature.

The single crystals of Bi2212 were grown using a self-flux method and then mechanically separated from the flux in $Al_2O_3$ or $ZrO_2$ crucibles [10]. The dimensions of the samples are typically 3×1×0.1 mm$^3$. The chemical composition of the Bi2212 phase corresponds to the formula $Bi_2Sr_{1.9}CaCu_{1.8}O_{8+x}$ as measured by energy dispersive X-ray fluorescence (EDAX). The crystallographic *a, b, c* values of the overdoped single crystals are of 5.41 Å, 5.50 Å and 30.81 Å, respectively. The $T_c$ value was determined by either dc-magnetization or by four-contacts method yielding $T_c$ = 87 - 90 K with the transition width $\Delta T_c$ ~ 1 K. The single crystals were checked out to assure that they are in the overdoped phase.

Experimental details of our break-junction technique can be found elsewhere [11]. Shortly, many break-junctions were prepared by gluing a sample with epoxy on a flexible insulating substrate and then were broken by bending the substrate with a differential screw at 14 - 18 K in a helium atmosphere. The electrical contacts were made by attaching gold wires to a crystal with silver paint. The sample resistance (with the contacts), $R_{sample}$, at room temperature varied from 5 Ω to about 1 kΩ, depending on the sample. The *I(V)* and *dI/dV(V)* tunneling characteristics were determined by the four-terminal method using a standard lock-in modulation technique. The tunneling spectra of our break junctions on Bi2212 single crystals exhibit below $T_c$ the characteristic features of typical tunneling spectra in Bi2212 [12,5,4].

Figure 1 shows a set of normalized tunneling conductance curves measured in superconductor-insulator-superconductor (SIS) between 14 K and 290 K on an overdoped Bi2212 single crystal with $T_c$ = 88.5 K. All curves, except the last one, show a gaplike structure. In the absence of generally accepted model for the pseudogap, we consider the presence of the gap-like features in tunneling spectra as a sign of the pseudogap (for more details, see Ref. 5). There is no sign indicating at what temperature the superconducting gap was closed. Across $T_c$, the superconducting tunneling spectra evolve *continuously* into a normal-state quasiparticle-gap structure which vanishes at 232 K < $T^*$ < 290 K but remains almost unchanged with temperature up to 232 K [13]. Such evolution of superconducting spectra into normal-state spectra across $T_c$ has been also observed in superconductor-insulator-normal metal (SIN) junctions [5]. Thus, it is clear that the pseudogap is related somehow to the superconductivity [5]. The spectra in Fig. 1 show the change in conductance from very low to high temperatures. It is difficult in such scale to see the differences between spectra above $T_c$. Therefore, Figure 2 shows tunneling spectra only above $T_c$, measured on another overdoped Bi2212 single crystal with $T_c$ = 88 K. In Fig. 2, one can see that there is a gap-like structure just above $T_c$ which disappears at some temperature $T_0$ much lower than $T^*$. Most pseudogap data observed in our study look similar to the data presented in Fig. 1, however, measurements on the best samples [14] show some gap-like features in tunneling spectra above $T_c$, which look similar to the gap-like structure shown in Fig. 2. Therefore, we believe that these features reflect intrinsic properties of overdoped Bi2212.

The spectra shown in Fig. 2 are asymmetrical about zero bias. Such behavior is typical for tunneling spectra obtained on the nonmetallic surfaces [15]. In fact, the contribution of the Cooper pairs to the pseudogap must not depend on the direction of bias. Thus, it is more convenient to consider only an even part of the conductivity, $G_e(V) \equiv [G(V) + G(-V)]/2$ [16]. Figure 3 shows the even part of the tunneling spectra presented in Fig. 2. In Fig. 3, one can clearly observe two humps at different temperatures, one of them disappears at $T_0 >$ 100 K, and the second one is traceable up to high temperatures. It is difficult to indicate exactly the origins of these humps, especially, the wide ones which are traceable to high temperatures. However, we will try just to determine characteristic temperatures from temperature dependencies of these humps, which are shown in Fig. 4. We use straight (dash) line in Fig. 4 in order to estimate the value of $T_0$ which is of the order of 110 - 115 K. The origin of this hump we discuss further. We concentrate now on the behavior of the curve with dots in Fig. 4, which correspond to the temperature dependence of wide humps. The fall of this curve with the increase of temperature between $T_c$ and $T_0$ supports an idea that there are some changes in the system. The indication of some changes which begin at ~ 210 K with the increase of temperature can be explained by the disappearance of local AF correlations [17]. It seems that, between $T_0$ and 210 K, there is no principal changes in the system. From Figs. 1, 2 and 3, the value of $T^*$ is of the order of 280 K. In Fig. 3, in addition to the two humps which have been discussed above, one can observe also a hump at 70 mV between 103 K and 122 K [18].

We discuss now the origin of the contribution to the pseudogap, which disappears at $T_0$ with the increase of temperature. This contribution can be explained by the disappearance of the CDW or/and SDW order(s), or incoherent Cooper pairs. From the common sense, the CDW or SDW order is likely to be appear above $T_c$ and not to disappear. Consequently, it is most likely that this contribution to the pseudogap is made by preformed pairs [8,9,5]. Tunneling spectroscopy is sensitive exclusively to the magnitude of the order parameter. Thus, if the humps originate from the presence of preformed pairs, then the magnitude of the gap-like structure reflects the average value of binding energy per particle [9]. However, our results alone cannot prove that this contribution originates from preformed pairs. We can argue that high-frequency conductivity measurements [8] showed that the incoherent Copper pairs remain well above $T_c$ in Bi2212, and it was predicted by the theory [17,19]. It is also in agreement with a MCS (Magnetic Coupling between Stripes) model proposed very recently [20,21]. There is also an alternative explanation for the presence of this contribution to the pseudogap, it is possible that this contribution originates from the presence of the Bi2223 phase with $T_c$ = 110 K [22]. Even in this case, our data obtained in SIS junctions support in general the conclusions made on the basis of tunneling data in SIN junctions, namely, that the pseudogap is directly related to superconductivity [5].

Recently, the pseudogap has been observed in Bi2212 with $T_c$ = 90 K (optimum doping) and $T_c$ = 84 K (slightly overdoped) by SIN tunneling technique [23]. The measured value of $T^*$ is equal to 300 K and 270 K, respectively. Thus, the value of $T^*$ ≈ 280 K found in our measurements for slightly overdoped single crystals is in an excellent agreement with the SIN tunneling measurements [23]. In slightly overdoped Bi2212, the pseudogap has been also observed in SIS tunneling measurements performed by break-junction technique [24]. The value of $T^*$ is found to be of the order of 190 K. However, the analysis of SIS spectra presented in Ref. 24 (see Fig. 4) shows that, in fact, the $T^*$ value is of the order of 290 - 300 K. In the literature, there is a clear discrepancy in the definition of $T^*$. For slightly overdoped Bi2212, the temperature 270 - 300 K corresponds most likely to the charge ordering, $T^*_{charge}$ ≈ 270 - 300 K [17]. The temperature $T^*$ ≈ 190 K found in Ref. 24 and the changes which were observed at ~ 210 K in our study (see Fig. 4) correspond most likely to the spin ordering, $T^*_{spin}$ ≈ 190 - 210 K [17]. Thus, there is a very good agreement among tunneling data presented in Refs. 23, 24 and the present work. What is interesting, in SIS tunneling measurements, the value of 116 K has been also discussed and assumed to be the local onset of superconductivity [24].

In summary, we presented direct measurements of the density-of-state by tunneling spectroscopy on slightly overdoped $Bi_2Sr_2CaCu_2O_{8-x}$ single crystals in a temperature range between 14 K and 290 K using the break-junction technique. The pseudogap which has been detected above $T_c$ appears at $T^*$ ≈ 280 K. The fine analysis of the spectra shows that there is a contribution to the pseudogap above $T_c$. We associate this contribution with the presence of incoherent Cooper pairs in Bi2212, which disappears approximately at 110 - 115 K. However, alternative explanations for the presence of this contribution are also possible. In all cases, there is no doubts that the pseudogap observed in our study is partially or entirely related to the superconductivity.

* * *

I thank R. Deltour for discussion. This work is supported by PAI 4/10.

look similar to the spectra shown in Fig. 2 are observed as rule in more "metallic" samples.

FIGURE CAPTIONS:

FIG. 1. Tunneling spectra measured in a SIS junction as a function of temperature on an overdoped Bi2212 single crystal. The conductance scale corresponds to the 290 K spectrum, the other spectra are offset vertically for clarity. The curves have been normalized at -150 mV (or nearest point).

FIG. 2. Tunneling spectra measured as a function of temperature on an overdoped Bi2212 single crystal. The conductance scale corresponds to the 290 K spectrum, the other spectra are offset vertically for clarity. The curves have been normalized at -200 mV (or nearest point).

FIG. 3. Even conductance of the spectra shown in Fig. 2 as a function of temperature.

FIG. 4. Temperature dependence of quasiparticle DOS shown in Fig. 3. The dashed line is a guide to the eye.

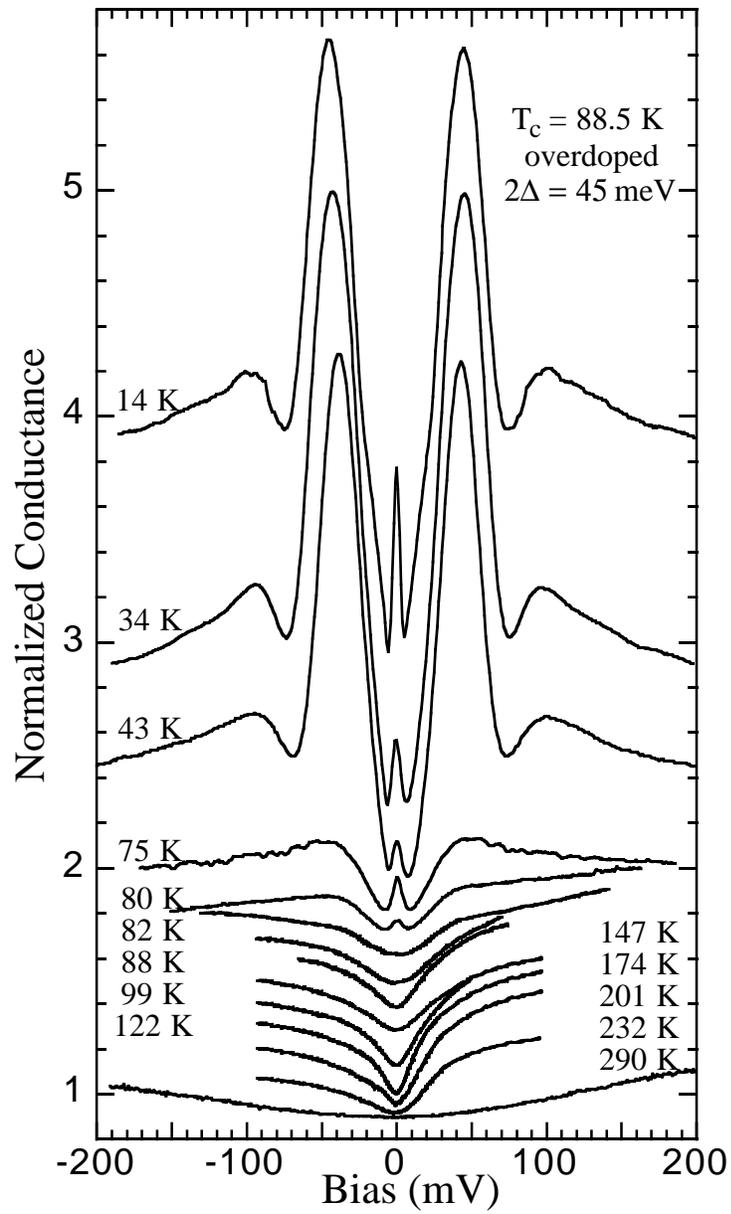

FIG. 1

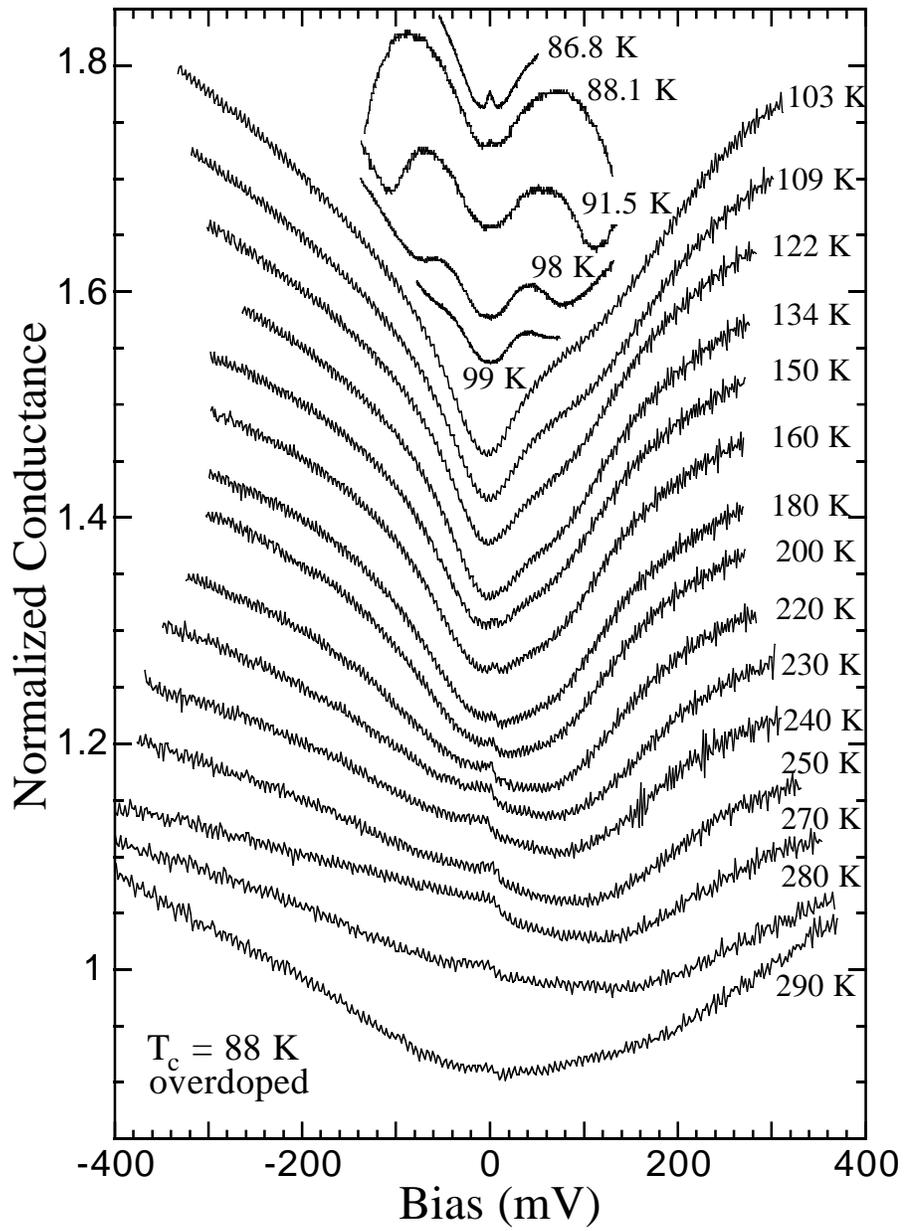

FIG. 2

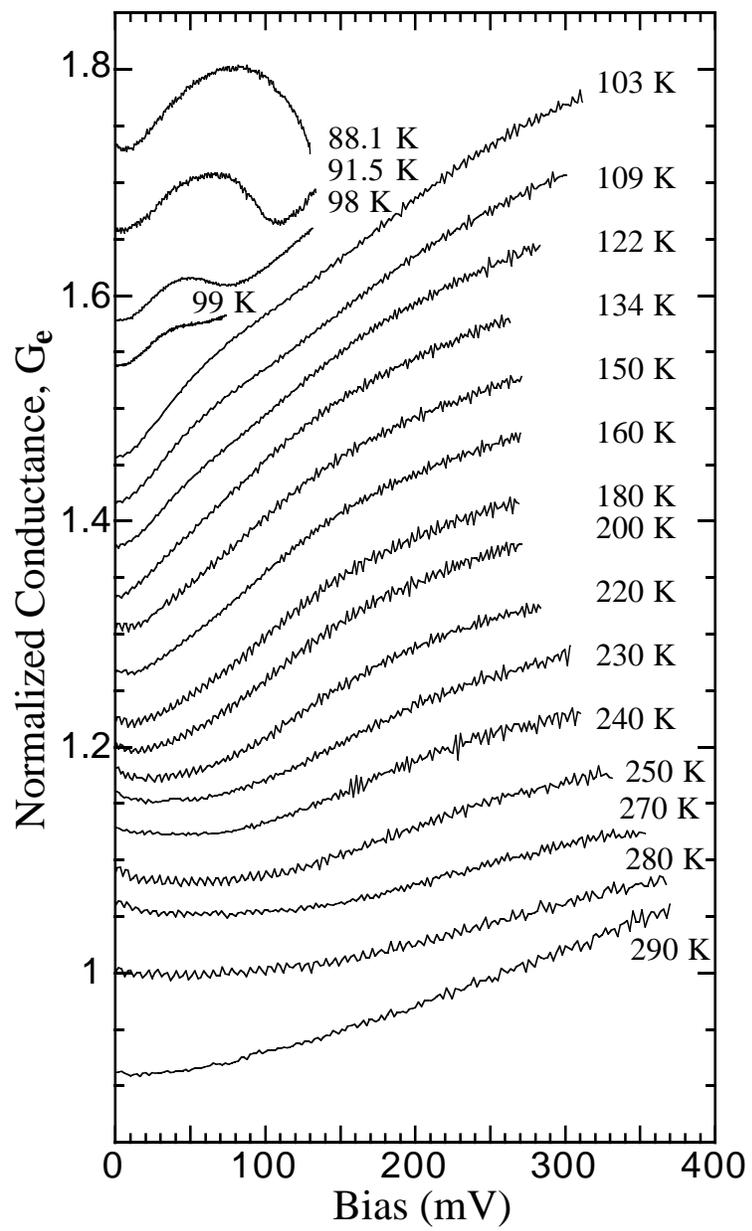

FIG. 3

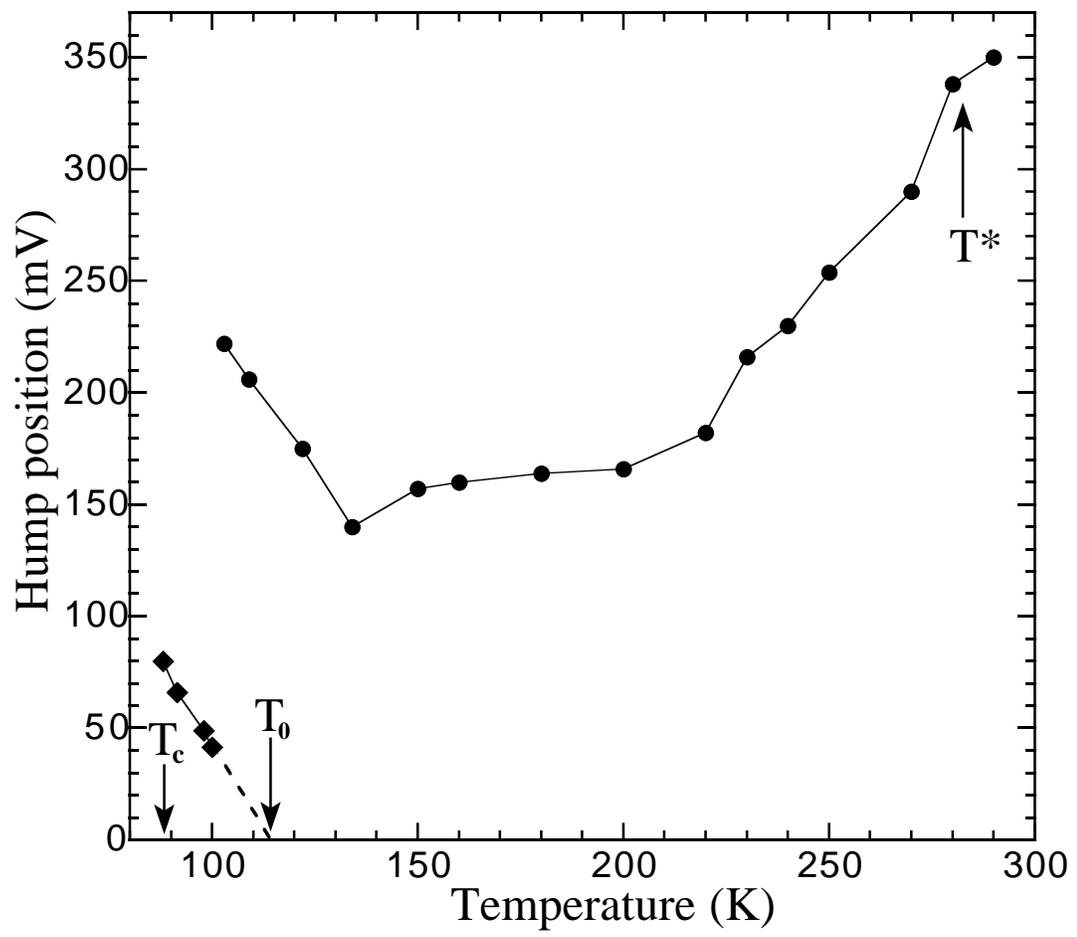

FIG. 4